\documentclass[reprint,twocolumn,pre,floatfix]{revtex4}
\usepackage{amsmath}
\usepackage{amsfonts}
\usepackage{amssymb}
\usepackage{graphicx}
\usepackage{wrapfig}
\usepackage{placeins}
\usepackage{color}
\usepackage{array}
\usepackage{dcolumn}

\newcommand\mfpt{\langle\tau(x)\rangle}
\newcommand\mfpty{\langle\tau(y)\rangle}
\newcommand\msd{\langle{X^2}\rangle}
\newcommand\MFPT{\text{MFPT}}
\newcommand\mean[1]{\langle{#1}\rangle}

\newcommand{\ropt}{r_{\text{opt}}}

\begin{document}
\pagestyle{empty}
\author{Oleg Semenov}
\email{olegsa@cs.unm.edu}
\author{David Mohr}
\email{dmohr@cs.unm.edu}
\author{Darko Stefanovic}
\email{darko@cs.unm.edu}
\thanks{to whom correspondence should be sent}
\affiliation{Department of Computer Science, University of New Mexico, MSC01 1130, 1 University of New Mexico, Albuquerque, NM 87131-0001}
\affiliation{Center for Biomedical Engineering, University of New Mexico, MSC01 1141, 1 University of New Mexico, Albuquerque, NM 87131-0001}

\title{First Passage Properties of Molecular Spiders}
\begin{abstract}
Molecular spiders are synthetic catalytic DNA-based nanoscale walkers.
We study the mean first passage time for abstract models of spiders moving on a finite two-dimensional lattice with various boundary conditions, and compare it with the mean first passage time of spiders moving on a one-dimensional track.
We evaluate by how much the slowdown on newly visited sites, owing to catalysis,
can improve the mean first passage time of spiders and show that in one dimension, when both ends of the track are an absorbing boundary, the performance gain is lower than in two dimensions, when the absorbing boundary is a circle; this persists even when the absorbing boundary is a single site.
 
\end{abstract}
\maketitle

\section{Introduction}
Natural molecular motors play an important role in biological processes that are critical for the functioning of living organisms; they are the source of most forms of motion in living beings~\cite{Schliwa:2003, Vale:2000, PhysBiolOfTheCell}.
In addition to naturally occurring molecular motors, several synthetic molecular motors have been designed~\cite{Yin:2004, Bath:2007, doi:10.1021/nl1037165, Wickham:2012, Shin:2004, Venkataraman07, Omabegho03042009, Delius:2010, ANIE:ANIE200504313}. 
Our work is inspired by a particular type of synthetic molecular motors---molecular spiders.

\begin{figure}[htbp]
    \includegraphics[width=0.35\textwidth]{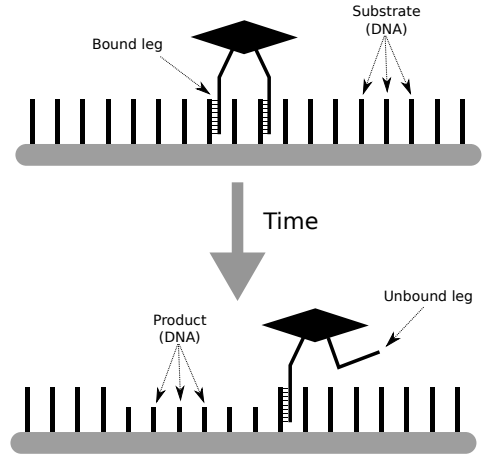}
    \caption{(Color online) 
    Molecular spider moves over a surface of single-stranded DNA substrates. It has several flexible deoxyribozyme legs. 
    When a leg detaches, it cleaves a substrate strand, turning it into a shorter product DNA strand. The leg can re-attach, but the bond will be weaker.}
    \label{fig:molecular-spider}
\end{figure}

Molecular spiders~\cite{Stojanovic:2006,Lund:2010} are synthetic nanoscale walkers which consist of a rigid, 
inert chemical body to which multiple flexible legs are attached. 
The legs are deoxyribozymes---enzymatic sequences of single-stranded DNA that 
can bind to and cleave complementary strands of a DNA substrate. When many such
substrates are attached to a surface, a leg can move between substrates,
cleaving them and leaving behind product DNA strands. Products can be revisited by a leg, but they cannot be cleaved again (Fig.~\ref{fig:molecular-spider}). 
The leg cleaves, and then detaches, more slowly from a substrate
than it detaches from a product.
The number of legs, and their lengths, can 
be varied, and this defines how a spider moves on the surface, i.e., its gait.

Mathematical models of molecular spiders at various levels of abstraction have been proposed and studied. Antal and collaborators 
introduced the first abstract model of molecular spiders, and studied the motion of a single spider on a one-dimensional track. 
They investigated the  movement of spiders with various numbers of legs and various gaits over products only~\cite{Krapivsky:2007a} and showed that such spiders are equivalent to a regular diffusion; the diffusion constants were computed for some gaits. 
Subsequently they introduced substrates and took into account that cleavage and detachment from substrates together take more time than the detachment from products~\cite{Krapivsky:2007b},
showing that this difference in residence time and the presence of multiple legs, together, bias a spider's
motion towards fresh substrates when it is on a boundary between substrates and products. This important property was also observed experimentally~\cite{Lund:2010}. In Ref.~\cite{spiderpaper:2011} we showed that spiders move superdiffusively for long periods of time. Samii et al. investigated various gaits and numbers of legs~\cite{PhysRevE.81.021106,PhysRevE.84.031111}, 
emphasizing the possibility of detachment from the track. In Refs.~\cite{spiderpaper:2011b, spiderpaper:2012} we studied the behavior of multiple spiders continuously released onto a 1D track. 
In a model with more physical detail~\cite{Olah:2012}, we showed that spiders can move against a force applied to the body. Models of spiders in two dimensions have also been studied.
In Ref.~\cite{wivace-12} we investigated how fast several spiders with various gaits can locate a small number of targets placed on a small fixed-size two-dimensional lattice. In Ref.~\cite{PhysRevE.85.061927} Antal and Krapivsky evaluated the diffusion constant and the amplitude describing the asymptotic behavior of the number of visited sites for a single spider with various gaits placed on an infinite square lattice. Analytical results regarding the asymptotic behaviors (limit theorems, transience, recurrence, and rate of escape) of spiders have been derived in Refs.~\cite{PSS:8474533, Ben-Ari:2011}. In Refs.~\cite{1742-5468-2007-11-P11015,Gallesco:2011} the behavior of spiders in random environments was studied. Rank et al. showed that several spiders, each placed on a separate 1D track and connected to a single cargo particle move it faster and remain superdiffusive longer than a single spider on a single 1D track~\cite{PhysRevE.87.032706}. 

Here we study first passage properties of an abstract theoretical model of molecular spiders. Our model is a direct extension of the model introduced in Refs.~\cite{Krapivsky:2007a,Krapivsky:2007b}. Although it is inspired by real molecular spiders, the model can also be applied to a wider class of random walkers that exhibit properties similar to spiders. Particularly, we investigate how the various  boundary conditions affect the spider's mean first passage time (\MFPT)
when it moves over finite one- and two-dimensional surfaces.

First passage properties of regular random walkers moving over discrete surfaces with various reflecting and absorbing boundaries have been extensively studied~\cite{PhysRevLett.95.260601, 1751-8121-44-2-025002}. 
Here we show that the difference between the time the legs spend on visited and on unvisited sites reduces the \MFPT{} of two-legged spiders in various surface settings, and increases the \MFPT{} of one-legged spiders (which behave like regular random walkers).

We start with the model of a two-legged spider that moves over a one-dimensional track with absorbing boundaries at both ends of the track. We found that for this surface the cleavage rate significantly affects the \MFPT{}, and for any track length there exists an optimal cleavage rate. 
Next, we study an extension of the 1D model to 2D, i.e., the mean first passage time of a two-legged spider to a circle, 
where the spider starts in the center. For this surface we determined that the cleavage rate gives the spider an even greater advantage over a regular random walker. The advantage persists even when the target is a single site, and thus is much harder to find. In this second 2D model the circle is a reflecting boundary, its center is an absorbing boundary, and the spider starts from various distances from the center.

\section{Model Details}
\label{sec:ak-model}
Our model is a modification of the model we used in Ref.~\cite{wivace-12}. It also can been seen as  a direct extension of the AK model of the spiders on a plane~\cite{PhysRevE.85.061927} that takes various boundary conditions into account.

\subsection{Motion and Surface}
In our model a single spider moves over finite one- or two-dimensional regular lattices. A spider has $k$ legs. It moves by detaching 
a leg
from its site on the lattice and reattaching it to a new site. Only one leg can detach at any given time, so a spider cannot detach all of its legs to leave the lattice. Each site can be occupied only by one leg at a time. There is a restriction on the maximum distance between any two legs $S$ (the gait), and each leg can move to one of the nearest neighboring sites (2 sites in 1D and 4 sites in 2D; a diagonal step is not allowed) with equal probability as long as the move does not violate one of the constraints above. Here we use only two types of spiders. First, a spider with $k=1$; this spider is equivalent to a regular random walker. The parameter $S$ does not affect this spider since it has only one leg. Second, a spider with $k=2$ and $S=2$; this spider is exactly the bipedal Euclidean spider with maximal separation $2$~\cite{PhysRevE.85.061927}. When such a spider is placed on a one-dimensional lattice, the model becomes equivalent to that of Ref.~\cite{spiderpaper:2011}. 

Two types of sites can be present on the lattice, substrate and product. A leg detaches at rate $1$ from a product, and at rate $r$ from a substrate, where normally $r\leq1$. Reattachment is instantaneous. When a leg leaves a substrate, that substrate is transformed into a product. Initially all sites are substrates; so any unvisited site is always a substrate, and a visited site is always a product. For $r<1$ the legs act differently when they are on visited versus unvisited sites, as described above. On the other hand, when $r=1$, legs act effectively the same whether they are on the products or substrates, and it becomes irrelevant if a site is visited (a product) or unvisited (a substrate). Thus, spiders with $r<1$ can be seen as having memory, since they react differently to visited and unvisited sites, and spiders with $r=1$ can be seen as having no memory, since they do not make this distinction.

\subsection{Boundaries and Starting Positions}
We study the first passage time of spiders moving over one-and two-dimensional regular lattices
with various boundary conditions and initial configurations. In each case we are interested when
the spider reaches the absorbing boundary.

The boundaries effectively make all our surfaces finite.
In one dimension we use a 1D track of length $2x$. The spider starts its movement 
from the middle of the track (the origin), and absorbing boundary sites are located at each
end of the track, i.e., each one is $x$ sites away from the origin. 

In two dimensions the surface is bounded by a circle of radius $x$.
Here we study three types of boundaries and initial spider positions.
First, the circle is an absorbing boundary, and the spider starts from the center of the circle.
Second, the circle is a reflective boundary, while the center is an 
absorbing boundary, and the spider can start from any site on the circle. Third,  we study the case when radius $x$ is fixed and the spider starts $y$ sites away 
from the target site at the center.

In all these settings $x$ effectively defines the size of the surface. And in all those settings, except the last, 
we study the dependence of \MFPT{} on $x$, i.e., $\mfpt$. In the last case we study the dependence of \MFPT{} on $y$, i.e., $\mfpty$.

\subsection{Meaningfulness of \MFPT{} in The Studied Settings}
A process to determine if \MFPT{} is a valid characteristic of the first passage behavior was given in Ref.~\cite{PhysRevE.86.031143}.
Similarly to that, we assess the meaningfulness of the mean first passage time in all studied settings. For every setting we estimate the distribution $P(\omega)$ of the random variable $\omega = \tau_1 / (\tau_1 + \tau_2)$, where $\tau_1$ and $\tau_2$ are first passage times of two independent spiders. Values of $\omega$ close to $1/2$ indicate that spiders act similarly in a particular setting. When $\omega$ is close to $0$ or $1$, the process is not uniform, and \MFPT{} is not a good measure of actual behavior. Distribution $P(\omega)$ can have three distinct shapes: unimodal bell-shaped, bimodal M-shaped, and plateau-like, almost uniform behavior. Bell-shaped form with a maximum at $\omega = 1/2$ indicates that \MFPT{} can be considered as a valid measure of the first passage times of individual spiders. M-shaped form with two peaks close to $0$ and $1$, and local minimum at $1/2$ indicates that \MFPT{} is not a good measure of the first passage time of individual spiders. The plateau-like shape with zero second derivative at $\omega = 1/2$ separates the two above cases. Just as in Ref.~\cite{PhysRevE.86.031143}, to quantify the shape of $P(\omega)$ we fit $P(\omega)$ to the model $\chi\omega^2 + c_1\omega + c_2$ for $0.05<\omega<0.95$. The sign of $\chi$ indicates the shape of $P(\omega)$. In cases when $\chi<0$, the distribution is bell-shaped; $\chi>0$ shows that the distribution is bimodal, M-shaped; and $\chi=0$ indicates that the distribution is almost uniform.

\subsection{Simulation}
Combining the states of the spider and the surface gives us a continuous-time
Markov process for our model. We use the Kinetic Monte
Carlo method~\cite{Bortz:1975} to simulate many trajectories of
the Markov process for every instance of the parameter set. In each case we record the first passage time to the absorbing boundary.

For Section~\ref{sec:1dtrack} we simulated $210$ different $r$ rates for up to a distance of $10000$ using $21504$ traces of the Markov process. For Section~\ref{sec:2dplanea} we simulated $100$ different $r$ rates for up to a distance of $1000$ using $20000$ traces. For Sections~\ref{sec:2dplaneb} we simulated $18$ different $r$ rates for up to a distance of $250$ using $20000$ traces.
For Section~\ref{sec:2dplanec} we simulated $15$ different $r$ rates for up to a distance of $100$ using $20000$ traces.

\section{Spider on a 1D Track}
\label{sec:1dtrack}
\subsection{Background: Transient Superdiffusivity of Spiders}
\label{sec:1d-msd-spider-results}

Antal and Krapivsky analytically obtain the mean time $\mean{T(n)}$ to visit $n$ sites in one dimension~\cite{Krapivsky:2007a,Krapivsky:2007b}.
Eq.~\ref{eq:T_n}
shows how $\mean{T(n)}$ depends on the value of $r$ for a two-legged spider ($k=2$).

\begin{equation}
\label{eq:T_n}
\mean{T(n)}=\frac{3}{2}\frac{1+r}{2+r}n^2 + \frac{1}{r} n.
\end{equation}

The parameter $r$ affects the leading asymptotic behavior of $\mean{T(n)}$. 
Eq.~\ref{eq:T_n_k1}, also from Ref.~\cite{Krapivsky:2007b}, gives $\mean{T(n)}$ for a one-legged spider

\begin{equation}
\label{eq:T_n_k1}
\mean{T(n)}=\frac{n(n-1)}{4} + \frac{n}{2r}.
\end{equation}

Eq.~\ref{eq:T_n_k1} demonstrates that for the one-legged spider ($k=1$), the leading term of $\mean{T(n)}$ does not depend on the parameter $r$. The $r$ only slows the one-legged spider down by increasing the sub-leading term. Eq.~\ref{eq:T_n_k1} and Eq.~\ref{eq:T_n} also show that in the absence of memory ($r=1$) the one-legged spider is faster than the two-legged spider. Thus these two properties, which separately make the walkers slower, surprisingly improve the performance of the spiders when combined together. This happens because difference in residence time between visited and unvisited sites biases multi-legged spiders towards unvisited sites when they find themseves on the boundary between visited and unvisited sites.

Using Kinetic Monte Carlo~\cite{Bortz:1975}  simulations of the Markov process 
we showed the unanticipated result that spiders of the AK model with $k=2$, $S=2$, and $r<1$
move superdiffusively over a significant span of time and distance
before eventually slowing down to move diffusively~\cite{spiderpaper:2011}.

Superdiffusive motion can be described using mean square displacement of a walker as a function of time. The mean squared displacement is given by Eq.~\ref{eq:msd}, where $d$ is the number of dimensions, and $D$ is an amplitude.

\begin{eqnarray}
\label{eq:msd}
\msd=2dD t^{\alpha} \quad
\begin{cases} \alpha=0 & \text{stationary}\\
0<\alpha<1 & \text{subdiffusive}\\
\alpha=1 & \text{diffusive}\\
1<\alpha<2 & \text{superdiffusive}\\
\alpha=2 & \text{ballistic or linear}
\end{cases}
\end{eqnarray}

From Eq.~\ref{eq:msd} we derive the condition for the walker to be superdiffusive at time $t$. The walker is moving \emph{instantaneously
superdiffusively}~\cite{Lacasta:2004} at a given time $t$ if
 
\begin{equation}
\alpha(t)=\frac{d(\ln{(\msd(t))})}{d(\ln{(t)})}>1.
\label{eq:alpha}
\end{equation}
 
In~\cite{spiderpaper:2011}, we showed that each spider process goes through three 
different phases
of motion defined by its value of $\alpha$. 
Initially spiders are at the origin, and must
wait for both legs to cleave a substrate before they start moving at
all. So when $t<1/r$ the process is essentially stationary (the \emph{initial phase}).  After the
spiders with $r<1$ take several steps, they show a sustained period
of superdiffusive motion over many decades in time (the
\emph{superdiffusive phase}).  Finally,
as time goes to infinity,
all spiders will approach ordinary diffusion
with $\alpha\approx 1$ (the \emph{diffusive phase});
the spiders mainly move over regions of previously
visited sites. which makes the value of $r$ less relevant.

\subsection{Mean First Passage Time}
\begin{figure}[htbp]
    \includegraphics[width=0.24\textwidth]{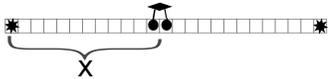} 
    \caption{Initial configuration of the track. All sites are initially substrates. Absorbing boundary is represented by stars.}
    \label{fig:1dstart}
\end{figure}

We measure the mean first passage time, $\mfpt$, where $x$ is the absolute distance of the walker from the origin on a one-dimensional track. At that point the walker is absorbed by the boundary. Fig.~\ref{fig:1dstart} shows the initial configuration of the track where the walker is positioned in the middle, and the absorbing boundaries are shown as stars. The length of the track is $2x$. 

According to the results of our numerical simulations the \MFPT{} of both one- and two-legged spiders is proportional to $x^2$. Eq.~\ref{eq:m_x_1d} shows the leading and sub-leading terms of $\mfpt$.

\begin{equation}
\label{eq:m_x_1d}
\mfpt \approx A_1(r)x^2+a_1(r)x.
\end{equation}

\begin{figure}[htbp]
	\includegraphics[width=0.48\textwidth]{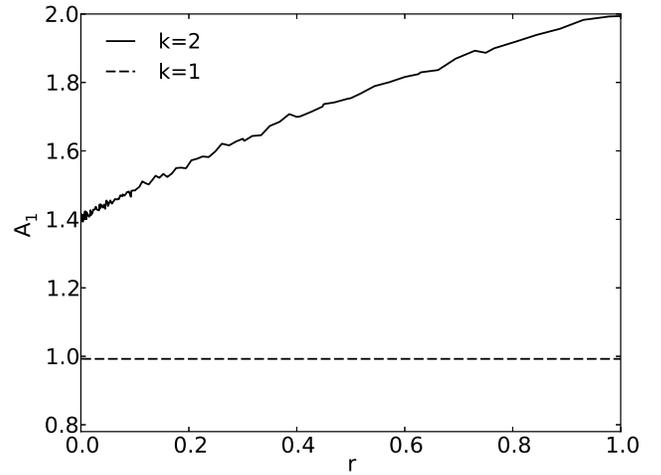}
	\caption{Amplitude $A_1$ from Eq.~\ref{eq:m_x_1d} as a function of the cleavage and detachment rate $r$. The data are for one-legged ($k=1$) and two-legged ($k=2$) spiders on a one-dimensional track.}
	\label{fig:A_1D}
\end{figure}

The amplitude $A_1$ of the leading term describes the asymptotic behavior of the \MFPT{}. For the one-legged spider we found that $A_1$ does not depend on $r$. For the two-legged spider $A_1$ increases with $r$, and approaches $2$ when $r=1$. As $r$  approaches zero $A_1 \approx 1.4$.  Fig.~\ref{fig:A_1D} shows the amplitude $A_1$ for one- and two-legged spiders for $182$ $r$ values less than $1$.

For both one and two-legged spiders the amplitude $a_1$ of the sub-leading term decreases monotonically and approaches $0$ in the absence of memory ($r=1$). The amplitudes $A_1$ and $a_1$ show that the parameter $r$  only increases the \MFPT{} of one-legged spiders. For two-legged spiders varying $r$ can decrease their \MFPT{}.

In Section~\ref{sec:1d-msd-spider-results} we recalled
that when $r<1$, the AK-model walkers go through three different regimes
of motion---the initial, superdiffusive, and the diffusive stage.
For lower $r$
values the initial slow period is longer than for higher $r$ values, but
subsequently the superdiffusive period is longer and faster.  For travel over shorter distances the initial period
is more important and thus larger $r$
values result in lower first passage times. For travel over longer distances the
superdiffusive period is important and smaller
$r$ values give better results. Thus for every particular distance
there is an optimal value of $r$ that minimizes the \MFPT{}.
For example, for distance 2000 the spider with $r=0.05$ is faster than 
the other (sampled) $r$ values; but for distances 4000 and longer 
the spider with $r=0.01$ is faster. 

\begin{figure}[htbp]
    \includegraphics[width=0.48\textwidth]{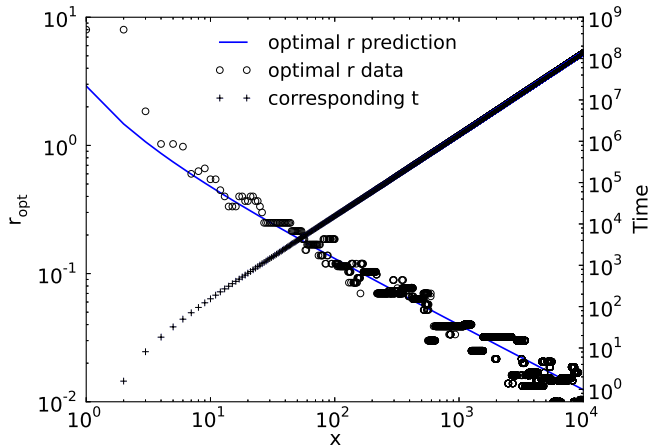}
    \caption{Optimal $r$ values ($r_{\text{opt}}(x)$) for various distances, and the corresponding $\mfpt$. Optimal $r$ values are obtained in two ways; first, by fitting the data for $\mfpt$ into the model of Eq.~\ref{eq:m_x_1d} and amplitudes $A_1$ and $a_1$ into the model of Eq.~\ref{eq:c_fit}, and second, by simulating $210$ values  of $r$ and choosing those that correspond to the lowest $\mfpt$. The estimation of $\ropt(x)$ shows that $\ropt(x) \sim x^{-1/2}$ .}
    \label{fig:1D-opt-r}
\end{figure}

We estimated $\ropt(x)$ for tracks of various lengths (up to 10000) through simulations of $210$ $r$ values. 
The results are shown in Fig.~\ref{fig:1D-opt-r}. The figure also shows the $\mfpt$ that corresponds to $\ropt(x)$ for each distance, i.e., the minimum $\mfpt$ achievable by varying the parameter $r$.
The optimum $r$ monotonically decreases with distance.
Smaller $r$ values create a stronger bias towards unvisited sites, but they make individual steps slower when the spider is on the boundary. Fig.~\ref{fig:1D-opt-r} shows that for better performance on shorter distances faster steps are more important than the bias towards unvisited sites, whereas for longer distances the bias dominates the \MFPT.

We also estimate $\ropt(x)$ by fitting. First, we assume that $A_1(r)$ and $a_1(r)$ have the functional form of Eq.~\ref{eq:c_fit}, and find the constants $c_1$ to $c_8$ by fitting Eq.~\ref{eq:c_fit} to the estimates of $A_1$ and $a_1$.

\begin{equation}
\label{eq:c_fit}
\begin{split}
A_1(r)=(c_1r+c_2)/(c_3r+c_4)\\
a_1(r)=(c_5r+c_6)/(c_7r+c_8)
\end{split}
\end{equation}

Then, we substitute the results into Eq.~\ref{eq:m_x_1d} and find when $\mfpt$ is minimized by extracting the derivative of $\mfpt$ with respect to $r$. The predicted $\ropt(x)$ is also drawn in Fig.~\ref{fig:1D-opt-r}. The derivation also shows that $\ropt(x) \sim x^{-1/2}$ (see the Appendix for details).

Our estimates of the indicator $\chi$ for various $r$ and $x$ values show that in the setting of the one-dimensional track $\chi$ is always negative, independent of the values of the parameters $x$ and $r$. Thus the \MFPT{} is a meaningful, reliable measure of the first passage time of spiders moving on a one-dimensional track.

\section{Spider on a 2D Plane with a Circular Absorbing Boundary}
\label{sec:2dplanea}
\begin{figure}[htbp]
    \includegraphics[width=0.24\textwidth]{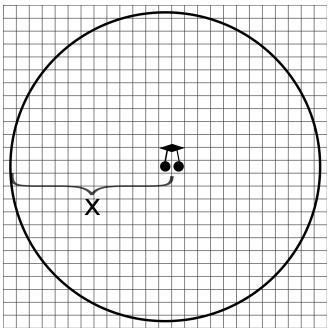}
    \caption{Initial configuration of the 2D surface. All sites are initially substrates. The spider starts at the center of the circle. The absorbing boundary is shown as a circle; as soon as either leg  crosses the circle the target is considered to be found, and the experiment stops. }
    \label{fig:2dstart}
\end{figure}

A direct extension of a one-dimensional track with length $2x$ to two dimensions is a set of sites bounded by a circle of radius $x$.
Fig.~\ref{fig:2dstart} shows the initial configuration of the surface with the spider positioned in the middle.

Using numerical simulations we found that, similarly to the spiders in one dimension, the \MFPT{} of both one- and two-legged spiders is proportional to $x^2$. However, the sub-leading terms are much closer to the leading terms, and therefore are more important for estimating the \MFPT{}. This implies that in two dimensions spiders (especially those with very small cleavage and detachment rate, $r<0.05$) approach the asymptotic behavior especially slowly. Eq.~\ref{eq:m_x_2d} shows the leading and sub-leading terms of $\mfpt$.

\begin{equation}
\label{eq:m_x_2d}
\mfpt \approx \left\{
 \begin{tabular}{ll}
 $A_2(r)\,x^2 + a_2(r)\,x^2/\ln t~~$ & : $k=2$\\
 $A_2(r)\,x^2 + a_2(r)\,x^2/(\ln t)^{0.88}$ &: $k=1$
 \end{tabular}
 \right.
\end{equation}

\begin{figure}[htbp]
	\includegraphics[width=0.48\textwidth]{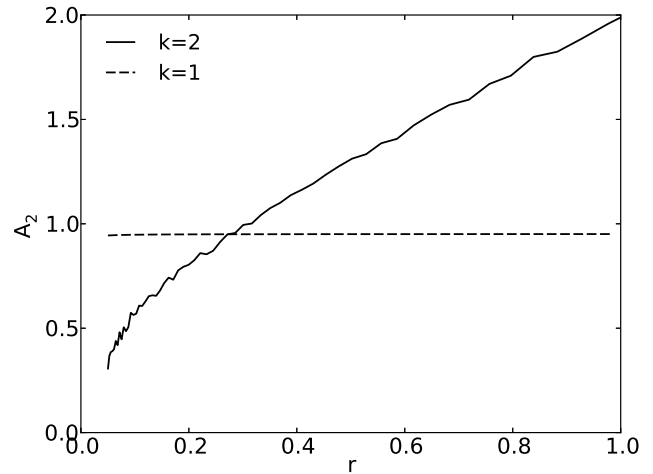}
	\caption{Amplitude $A_2$ from Eq.~\ref{eq:m_x_2d} as a function of the cleavage and detachment rate $r$. The data is shown for the one-legged ($k=1$) and two-legged ($k=2$) spiders moving over two-dimensional surface with a circular absorbing boundary.}
	\label{fig:A_2D}
\end{figure}

The correlation $(\ln t)^{-0.88}$ for the one-legged spider is unusual and slower than for the two-legged spider. Somewhat similar effects were observed in Ref.~\cite{PhysRevE.85.061927} in the estimation of the mean squared displacement. As in one dimension, the amplitude $A_2$ of the leading term does not depend on $r$ for the one-legged spider and is proportional to $r$ for the two-legged spider. The sub-leading term's amplitude $a_2$ monotonically decreases with $r$ in both cases. Fig.~\ref{fig:A_2D} shows the amplitude $A_2$ for the one- and two-legged spiders for $60$ $r$ values that are less than $1$.

For one-legged spiders, as in one dimension, lower $r$ values only increase the \MFPT{}. The comparison of Figs.~\ref{fig:A_2D} and~\ref{fig:A_1D} shows that for two-legged spiders $A_2$ is affected more strongly by $r$ than $A_1$. The $A_2$ of the two-legged spider in two dimensions even intersects the $A_2$ of the one-legged spider. The $A_2$ starts at $2$ when $r=1$ and decreases towards $\approx 0.5$ as $r$ approaches zero.
 
Similar to the one-dimensional case, spiders with $r<1$ start more slowly than the no-memory spider with $r=1$; then they move faster, and finally they slow down 
and approach regular diffusion. But in contrast to spiders on a 1D track, the $r$ values that correspond to fastest times are higher, and the \MFPT{} of spiders with very small $r$ values (less than $0.1$) approaches the \MFPT{} of spiders with $r=1$ very slowly. However, the transition towards the diffusive stage happens more slowly compared with 1D.

\begin{figure}[htbp]
    \includegraphics[width=0.48\textwidth]{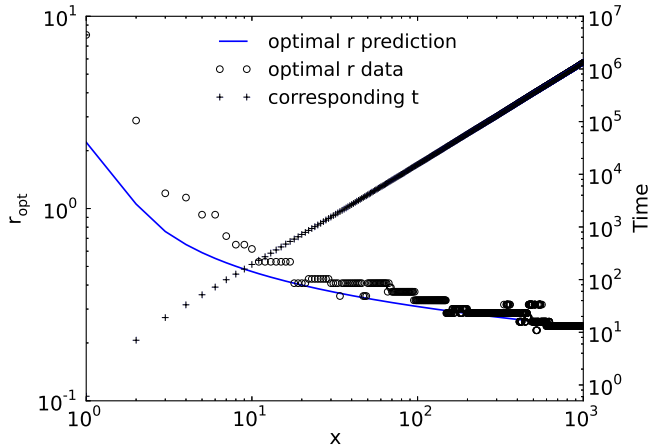}
    \caption{Optimal $r$ values ($\ropt(x)$) for various radii. Optimal $r$ values are obtained in two ways; first, by fitting the data for $\mfpt$ into the model of Eq.~\ref{eq:m_x_2d} and amplitudes $A_2$ and $a_2$ into the model of Eq.~\ref{eq:c_fit}, and second, by simulating $100$ values of $r$ and choosing those that correspond to the lowest $\mfpt$.}
    \label{fig:2D-opt-r}
\end{figure}

We estimated $\ropt(x)$ for circles of various sizes (up to 1000) through simulations of $100$ $r$ values, choosing the fastest ones for each distance. Comparison of the results is shown in Fig.~\ref{fig:2D-opt-r}. The figure also shows the fastest $\mfpt$ that corresponds to $\ropt(x)$ for each distance. Similarly to the 1D case, we also estimate $\ropt(x)$ by fitting. 

Our estimates of the indicator $\chi$ for various $r$ and $x$ values show that in the setting of the two-dimensional circle with absorbing boundary at the perimeter $\chi$ is always negative, and this does not change with the values of the parameters $x$ and $r$. Just as in 1D, the \MFPT{} is a meaningful measure of the first passage time of the individual spiders moving on a two-dimensional lattice.

\section{Spider on a 2D plane with a circular reflecting boundary and an absorbing boundary in the center}

When we change the contour of the 2D surface to be a circular reflecting boundary instead of an absorbing one, and place a single target site in the center, spiders with memory ($r<1$) still have an advantage over spiders without memory ($r=1$). We consider two cases: (1) when the radius $x$ of the circle is variable, and spider starts from any point on the contour, and (2) when the radius $x$ is fixed, and the distance $x$ between the starting position and the target is variable. Since both cases are circularly symmetric, all starting positions with the same distance from the center are equivalent. In both cases we study how $\mfpt$ is affected by the parameter $r$, and how it is affected by $x$ in (1) and $y$ in (2).

\subsection{2D Circle of Variable Radius With Target in the Middle and Spider Starting from the Boundary}
\label{sec:2dplaneb}

\begin{figure}[htbp]
    \includegraphics[width=0.24\textwidth]{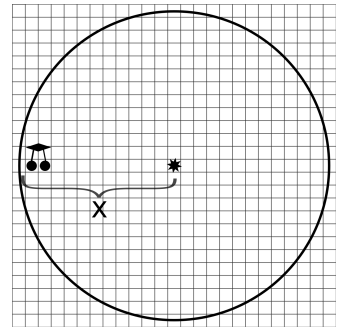}
    \caption{Initial configuration of the surface. All sites are initially substrates. The absorbing boundary is shown as a star. The spider starts at the periphery. The circle is a reflecting boundary and thus steps outside of the circle are not allowed.}
    \label{fig:2dcstart}
\end{figure}

Fig.~\ref{fig:2dcstart} shows the initial configuration of the surface where the spider is positioned on the contour and the single target site (absorbing boundary) is shown as a star. Since the reflecting boundary is a circle and the absorbing boundary is a single site in the center, all starting positions of the same distance from the target site are equivalent. In our simulations we pick a fixed position on the contour as a starting position for the spider. This position is shown as two small solid circles in Fig.~\ref{fig:2dcstart}.

We measure the first passage time of the walker from the contour of the surface to its center for various radii $x$. 

Interestingly, the \MFPT{} asymptotically grows slightly faster than $x^2$ . The sub-leading term also grows faster than in the circular absorbing boundary case. Eq.~\ref{eq:m_x_2d_rev} shows the leading and sub-leading terms of $\mfpt$.

\begin{equation}
\label{eq:m_x_2d_rev}
\mfpt \sim B_2(r)x^2 + b_2(r)x^{1.5}.
\end{equation}

\begin{figure}[htbp]
	\includegraphics[width=0.48\textwidth]{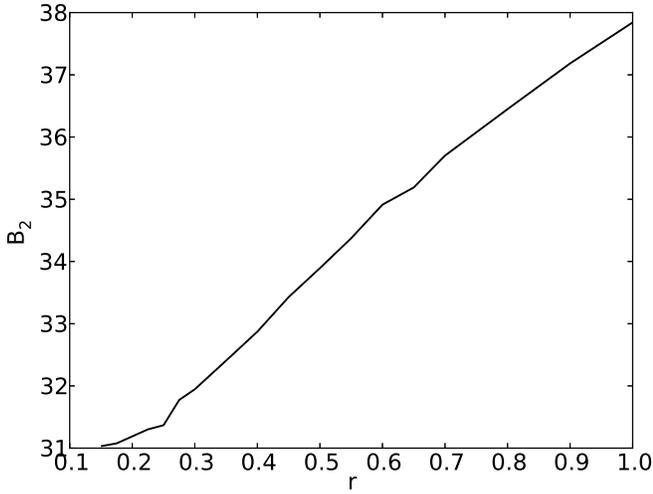}
	\caption{Amplitude $B_2$ from the Eq.~\ref{eq:m_x_2d_rev} as a function of the cleavage and detachment rate $r$. The data are shown for two-legged ($k=2$) spiders moving over a two-dimensional surface with a circular reflecting boundary. The single site in the middle of the circle is an absorbing boundary.}
	\label{fig:A_cirVarR2D}
\end{figure}

Fig.~\ref{fig:A_cirVarR2D} shows the amplitude $B_2$ for the two-legged spiders for $20$ different $r$ values between $0.1$ and $1.0$.

Values of $r$ smaller than $1$ give the spider an advantage, similar to the 2D configuration discussed above, but now even for shorter radii $x$. 
The typical time-line of this process can be characterized as follows. First, the spider starts moving, it eventually visits many sites without finding the target site, and leaves behind many smaller regions of substrates. Next, the spider starts to move only over visited sites more often, and eventually encounters regions of substrates of various shapes and sizes. At this stage, the spider with $r=1$ will not be affected by the substrate regions, and will move just as if they were visited sites. On the other hand, spiders with $r<1$ will become biased to stay on the substrates and explore those regions more thoroughly; this will increase their chances of finding the target site, since it must be in one of those unvisited areas. This scenario can explain how spiders with $r<1$ gain an advantage over spiders with $r=1$  on this surface. 

\begin{figure}[htbp]
    \includegraphics[width=0.48\textwidth]{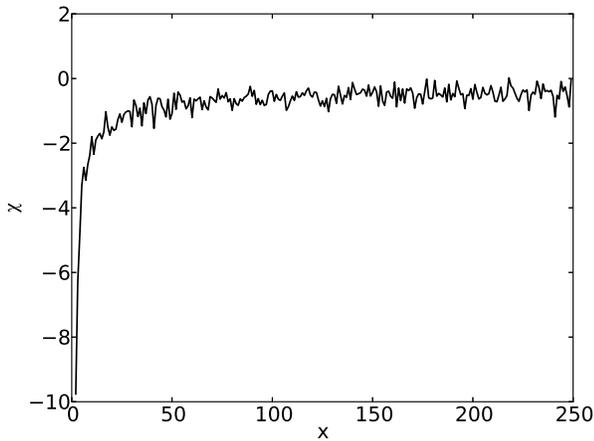}
    \caption{Indicator $\chi$ shows how well the \MFPT{} can describe the first passage times of the individual spiders to a single target. The surface is two-dimensional with a circular reflecting boundary of radius $x$, and spider starts from the contour.}
    \label{fig:Chi_cirVarR2D}
\end{figure}

The indicator $\chi$, in this setting, grows with the parameter $x$, and for $x\gtrapprox17$ $\chi$ becomes close to $0$. As in the previously described settings, $\chi$ does not depend on the parameter $r$. Fig.~\ref{fig:Chi_cirVarR2D} shows the dependence of $\chi$ on $x$ for $r=0.5$. 
For small surfaces (small values of $x$), the $\chi$ is negative, and thus in those cases \MFPT{} is a good measure of the first passage time of individual spiders. However, for larger $x$, close to $0$ values of $\chi$ indicate that the shape of the distribution $P(\omega)$ is close to uniform, and thus the \MFPT{} is not as good a measure of individual behavior as it is in the settings when the spider searches for the contour. It also shows that the possible paths that the spider can take to locate a single target are more diverse than paths that lead to the contour.

\subsection{2D Circle of Fixed Radius With Target in the Middle and Spider Starting From Various Distances}
\label{sec:2dplanec}

\begin{figure}[htbp]
    \includegraphics[width=0.24\textwidth]{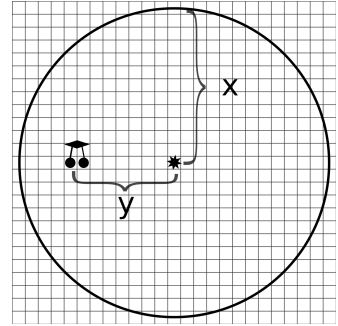}
    \caption{Initial configuration of the surface. All sites are initially substrates. The absorbing boundary is shown as a star. The spider starts $y$ sites away from the center. A circle of constant radius $x=100$ is a reflecting boundary and thus steps outside of the circle are not allowed.}
    \label{fig:2dxcstart}
\end{figure}

Fig.~\ref{fig:2dxcstart} shows the initial configuration of the surface where the walker is positioned at a distance $y$ from the target site, and the radius $x$ is set to 100 and remains constant. 

\begin{figure}[htbp]
    \includegraphics[width=0.48\textwidth]{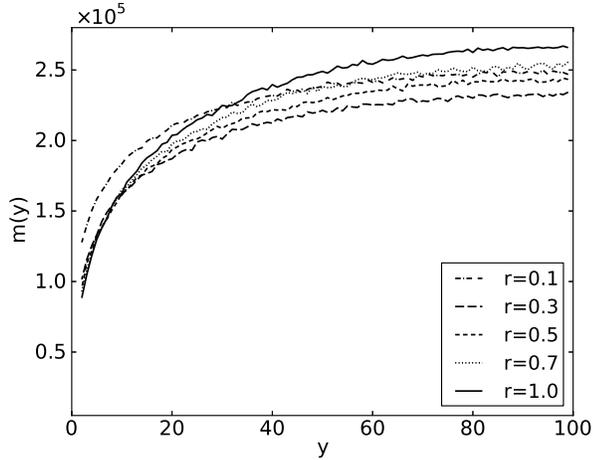}
    \caption{Mean first passage time to a single point in the center of the circle with a fixed radius. The boundary is reflecting, and spider starts at various distances from the target.}
    \label{fig:k2m01sm02_cir_sng__allr}
\end{figure}

We measure the first passage time of the walker from the contour of the surface to its center for various distances $y$. Fig.~\ref{fig:k2m01sm02_cir_sng__allr} shows a plot of the \MFPT{}. 

The shape of the curves is asymptotically logarithmic; this shows that the initial position of the spider does not significantly affect $\mfpty$ when $x$ is fixed. Even if the spider starts closer to the target, there are too many possible paths to the target in 2D, of which one is randomly chosen. Many of them are very long and initially lead the spider far away from the target. 

\begin{figure}[htbp]
    \includegraphics[width=0.48\textwidth]{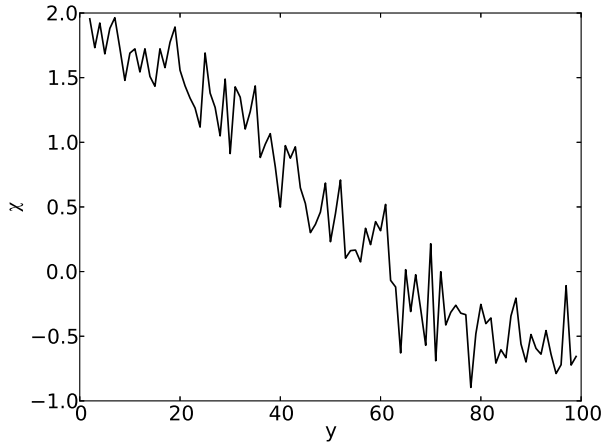}
    \caption{Indicator $\chi$ shows how well the \MFPT{} can describe the first passage times of the individual spiders to a single target. The surface is two-dimensional with a circular reflecting boundary of a fixed radius $x=100$, and spider starts at the distance $y$ from the target site.}
    \label{fig:Chi_cirVarX2D}
\end{figure}

The indicator $\chi$, in this setting, is positive for $y\lessapprox80$ and decreases with the parameter $y$. For $y\gtrapprox70$ as the surface configuration approaches the configuration discussed in the Section~\ref{sec:2dplaneb}, and $\chi$ becomes negative, but as in the case of the Section~\ref{sec:2dplaneb}, it still remains close to $0$. Fig.~\ref{fig:Chi_cirVarX2D} shows the dependence of $\chi$ on $y$ for $r=0.5$. 
The positive values of $\chi$ for the smaller $y$ values indicate that \MFPT{} is not a good measure of the first passage times of the individual spiders, and paths that lead the spider to a target are very diverse and vary significantly in their lengths.

\section{Comparison of Spider Performance in The Studied Settings}
\begin{figure}[htbp]
    \includegraphics[width=0.48\textwidth]{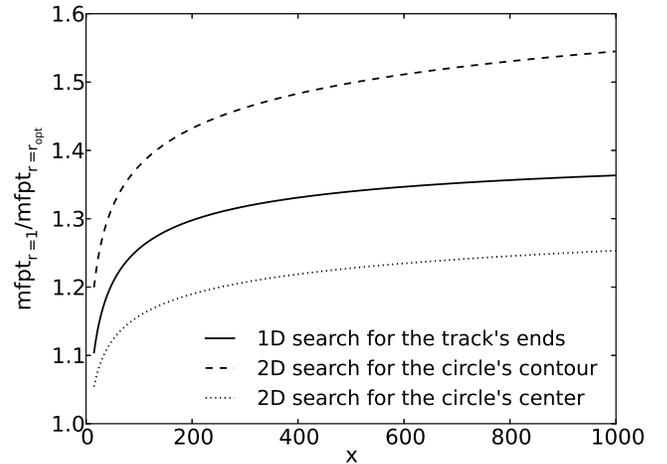}
    \caption{Ratio between $\mfpt$ of the spiders with $r=1.0$ and $r=r_\text{opt}$.}
    \label{fig:1D2D-eff}
\end{figure}

It is interesting to compare the advantage spiders with $r<1$ (i.e., with memory) enjoy over those with $r=1$ (i.e., without memory) in the described 1D and 2D settings. 
We compute the ratio of $\mfpt$ for spiders with $r=1$ and $\mfpt$ for spiders with $r=\ropt(x)$. 
This ratio is plotted in Fig.~\ref{fig:1D2D-eff} against the surface size for 1D and 2D.

\begin{figure}[htbp]
    \includegraphics[width=0.48\textwidth]{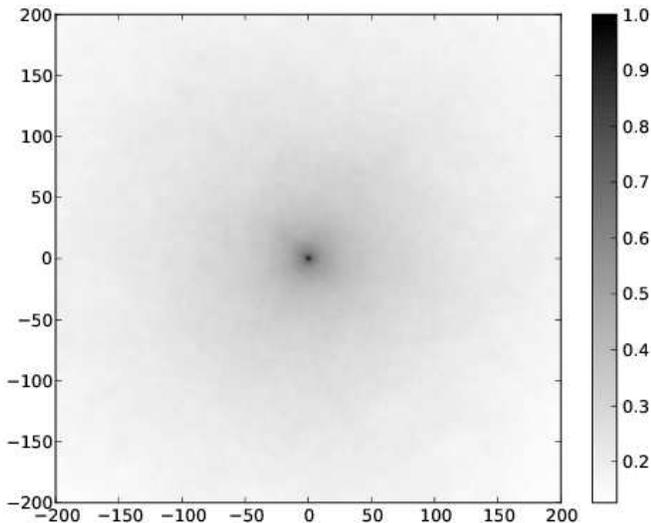}
    \caption{Average density of products when spider with $r=0.25$ and $k=2$ is at distance 1000.}
    \label{fig:2Dprod_den}
\end{figure}

The plot shows that the cleavage rate $r$ gives spiders more advantage on a 2D plane searching for a circle than on a 1D strip searching for its ends. This advantage can be attributed to the amount of substrates that spiders leave behind as they move away from the origin. In 1D, spiders do not leave any substrates behind, so there are no substrates between the left and the right ends of the sea of products, and when a spider moves back towards the origin there are no substrates to bias it towards the boundary. In 2D, the shape of the product sea can be very complicated and there can be many substrates left behind. When a spider moves backward it has a high probability to still encounter substrates, which can bias it towards the boundary. The higher $\ropt(x)$ values in 2D can be attributed to the direction of the emergent bias when the spider is on the border between visited and unvisited sites. In 1D, the border is simple, and its shape remains the same over time. It is defined by the two closest unvisited sites to the origin on the right and on the left side. As a result, when the spider with rate $r<1$ is on the border, it is always biased in the desired direction---away from the origin. In 2D, the shape of the border can be much more complex and is even not necessarily connected. This type of border leads to a much weaker bias towards the edge of the surface. In many cases the spider is not biased directly to the edge in the direction of the shortest path, and sometimes the spider even is biased back towards the circle's center. Despite that, the greater amount of substrates accessible by the spider in 2D overcomes the weaker bias and makes spiders more efficient at finding the absorbing boundary. Fig.~\ref{fig:2Dprod_den} shows the average density of products; the spider has a high probability of encountering substrates when it turns back towards the origin. 

\section{Discussion}

Our simulations show that the $\MFPT$ of two-legged spiders depends strongly on the kinetic parameter $r$ in all studied cases.
The $\MFPT$ is lower for $r<1$ (i.e., with memory) than for $r=1$ (i.e., without memory) in the one-dimensional case when the spider is searching for the ends of a one-dimensional track.
For one-legged spiders the parameter $r$ does not affect leading asymptotic behavior, however it slows them down by increasing the constant of the sub-leading term.
In the extension of this problem to two dimensions, when the spider is searching for the contour of a circle  from its center, the advantage of having $r<1$ is even more significant, despite the less effective bias provided by the shape of the leftover substrates. Here the bias provided by the substrates does not always direct the spider towards the absorbing boundary. In contrast, on a 1D track, the substrates always bias the spider towards the closest end when one of the legs is attached to them. The disadvantage in 2D is overcome by the greater amount of substrates that are accessible to spiders. In 1D, the spider (of the studied gait) does not leave any substrates behind when it progresses towards the ends of the track. In 2D, the shape of the sea of products is complex, and many substrates are left behind. Those substrates can bias the spider towards the absorbing boundary when it starts to turn back towards the origin.

When we reverse boundaries in 2D we make the absorbing boundary the single site in the center of the circle, and start the spider from the contour, the parameter $r$ can still improve the \MFPT{}. It is difficult to find the single target site in this scenario, and the \MFPT{} increases significantly for all types of walkers. However, it is plausible that when most of the surface is explored, spiders with $r<1$ are more likely to stick to small remaining islands of substrates (i.e., unvisited sites). The presence of these islands increases the probability that spiders with $r<1$ will find the target site, as opposed to a spider with $r=1$, which would not react to the presence of those islands. This is a likely explanation for the  greater importance of $r$ for this boundary condition.

Despite the varying importance of the parameter $r$ in the studied scenarios, in all cases there exists an optimal value of $r$ that minimizes the first passage time to the absorbing boundary.
To the extent that catalysis is accessible as a design parameter of molecular spider assemblies, the results provide a way to
optimize spider system performance in various target search scenarios.

\begin{acknowledgments}
The authors would like to thank Paul Krapivsky for detailed
discussions concerning their model and analysis.  This material is based upon work supported by the National Science Foundation under grants 0829896 and 1028238. We would also like to thank NVIDIA Corporation for a hardware gift that made possible some of the simulations.
\end{acknowledgments}

\bigskip

\appendix

\section{Optimum value of $r$}
First, we find the derivative of the assumed form pf $\mfpt$ (Eqs.~\ref{eq:m_x_1d} and \ref{eq:optrDir}) with respect to $r$.

\begin{align}
\label{eq:optrDir}
\frac{d\mfpt}{dr} = \frac{c_1(c_3r+c_4)-c_3(c_1r+c_2)}{(c_3r+c_4)^2}x^2 \notag\\
+ \frac{c_5(c_7r+c_8)-c_7(c_5r+c_6)}{(c_7r+c_8)^2}x\notag\\
=\frac{c_1c_4-c_3c_2}{(c_3r+c_4)^2}x^2 + \frac{c_5c_8-c_7c_6}{(c_7r+c_8)^2}x
\end{align}

The fitting of Eq.~\ref{eq:c_fit} to the estimates of $A_1$ and $a_1$ yields  estimates of constants $c_1$ to $c_8$. Next, we substitute the constants into Eq.~\ref{eq:optrDir} and find when the derivative is zero.

\begin{align}
\label{eq:optrDirzero}
\frac{3.29r+3.01}{r+2.15}x + \frac{-4.35r+1.34}{r} = 0\notag\\
r = \frac{5.75+\sqrt{4.05x-1.34}}{8.09x-2.68}\notag
\end{align}

\bibliographystyle{ieeetr}
\bibliography{spiders_fpt}

\end{document}